\documentclass[11pt]{article}

\usepackage{amsthm,amsmath,amssymb,amsfonts,xspace,graphics,enumerate}
\usepackage{color}
\usepackage{authblk}
\usepackage{cite}

\usepackage{graphicx}
\usepackage{array,rotating}
\usepackage{amsmath}
\usepackage{amsfonts}
\usepackage{lscape}
\usepackage{multirow}
\usepackage{booktabs}
\usepackage{mdwlist}
\usepackage{enumitem} 
\setlist[itemize]{noitemsep, topsep=0.5pt}

\usepackage{bm} 
\usepackage{mathtools} 
\usepackage[table]{xcolor}
\usepackage{url} 
\usepackage{wrapfig}

\usepackage{soul}

\usepackage{float}

\usepackage{natbib}
\usepackage{comment}
\usepackage{fullpage} 
\usepackage{graphicx}
\usepackage[font=small]{caption}
\usepackage{subcaption}
\usepackage{amssymb}
\usepackage{amsmath}
\usepackage{graphics}
\usepackage{bm}
\usepackage{color}
\usepackage{amsthm}
\usepackage{enumerate}
\usepackage{amsfonts}
\usepackage{bbm}
\usepackage[toc]{appendix}
\usepackage{enumitem}
\usepackage{xr}
\usepackage{changes}

\usepackage[compact]{titlesec}
\titlespacing{\section}{0pt}{2ex}{1ex}
\titlespacing{\subsection}{0pt}{1ex}{0ex}
\titlespacing{\subsubsection}{0pt}{0.5ex}{0ex}

\usepackage{tikz}
\usepackage[CJKbookmarks=true,
bookmarksnumbered=true,
bookmarksopen=true,
colorlinks=true,
citecolor=blue,
linkcolor=blue,
anchorcolor=blue,
urlcolor=blue]{hyperref}
\usepackage[ruled, vlined, lined, commentsnumbered,linesnumbered]{algorithm2e}

\DeclareMathOperator*{\argmin}{arg\,min}
\DeclareMathOperator*{\argmax}{arg\,max}

\newcommand{\bzero}{{\bf 0}}

\newcommand{\bA}{{\bf A}}

\newcommand{\bC}{{\bf C}}

\newcommand{\bba}{{\bf a}}
\newcommand{\bc}{{\bf c}}

\newcommand{\bh}{{\bf h}}
\newcommand{\bg}{{\bf g}}

\newcommand{\bw}{{\bf w}}

\newcommand{\gammam}{\mbox{\boldmath $\gamma$}}
\newcommand{\betam}{\mbox{\boldmath $\beta$}}

\newcommand{\Sigmam}{\mbox{\boldmath $\Sigma$}}

\newcommand{\Gamm}{\mbox{\boldmath $\Gamma$}}

\newcommand{\hmu}{{\hat{\mu}}}

\newcommand{\mT}{{\mathcal{T}}}
\newcommand{\mbR}{{\mathbb{R}}}

\newcommand{\htheta}{{\hat{\theta}}}
\newcommand{\hw}{{\hat{w}}}
\newcommand{\hV}{{\hat{V}}}

\newcommand{\hgammam}{\hat{\mbox{\boldmath $\gamma$}}}

\newtheorem{defi}{Definition}

\newtheorem{rmk}{Remark}
\newtheorem{asmp}{Assumption}

\begin{document}

\title{Functional Data Analysis with Causation in Observational Studies: Covariate Balancing Functional Propensity Score for Functional Treatments}

\author[1]{Xiaoke Zhang}
\author[1]{Wu Xue}
\author[2]{Qiyue Wang}
\affil[1]{Department of Statistics, George Washington University}
\affil[2]{Department of Computer Science, George Washington University}

\maketitle

\begin{abstract}
Functional data analysis, which handles data arising from curves, surfaces, volumes, manifolds and beyond in a variety of scientific fields, is a rapidly developing area in modern statistics and data science in the recent decades. The effect of a functional variable on an outcome is an essential theme in functional data analysis, but a majority of related studies are restricted to correlational effects rather than causal effects. This paper makes the first attempt to study the causal effect of a functional variable as a treatment in observational studies. 
Despite the lack of a probability density function for the functional treatment, the propensity score is properly defined in terms of a multivariate substitute. Two covariate balancing methods are proposed to estimate the propensity score, which minimize the correlation between the treatment and covariates. The appealing performance of the proposed method in both covariate balance and causal effect estimation is demonstrated by a simulation study. The proposed method is applied to study the causal effect of body shape on human visceral adipose tissue.
\end{abstract}


\linespread{1.5}


%
%
%
%
%

\section{Introduction} \label{sec:intro}

Functional data, which arise from curves, surfaces, volumes, manifolds and beyond, are increasingly common in a variety of fields due to recent technological innovations in data collection, data storage and scientific computing. The broad availability of functional datasets and immediate analytic demand have stimulated the rapid development of functional data analysis (FDA) in the past few decades and enhanced its importance in modern statistics and data science. For systematic introductions to FDA and select topics, readers may refer to
representative monographs \citep[e.g.,][]{Bosq00, RamsS05, FerrV06, HorvK12, Zhan13, HsinE15, KokoR17}. 


To study the association between a functional predictor and a scalar response, which is a classic topic in FDA, functional regression is the most popular class of approaches \citep[see survey papers, e.g., by][]{Morr15, ReisGS17}. 
However, functional regression can only reveal the correlation between the response and functional predictor, although the causal relationship between those is of primary interest in certain scientific studies. Despite its intellectual merit and practical impact, the research on causal inference in FDA has been inadequate. Among very few existing works, almost all of them focus on either clinical trials or functional variables as covariates \citep[e.g.,][]{Lind12, McKeQ14, CiarPO15, CiarPO18, 
ZhaoL19, MiaoXZ20}. 
In contrast, 
we consider a functional variable
as a treatment in observational studies. 



In the literature on causal inference for observational studies, the causal effect estimation of a binary treatment is a classic topic and has been intensively studied \citep[see reviews, e.g., by][]{Imbe04, 
Stua10, DingL18, YaoCL20}. In the past few decades, an increasing number of papers have appeared on multi-level categorical treatments 
or continuous treatments \citep[e.g.,][]{Imbe00, RobiHB00, HiraI04, ImaivD04, ZhuCG15, YangIC16, KennMM17, LopeG17, 
FongHI18, LiL19}. 
Recently, treatments of more complex forms have gradually received more attentions, such as multidimensional categorical treatments \citep[e.g.,][]{DAmo19, 
WangB19} 
and matrix treatments \citep[e.g.,][]{YuWK20} among others. To the best of our knowledge, this paper is the first to handle functional treatments in observational data.

\begin{figure}[H]
\centering
\includegraphics[width=4cm]{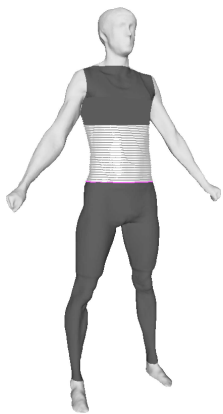}
   \caption{Body circumference over the trunk and two legs of a person sampled at $128$ equidistant levels from neck to ankle.   
   The shadeless portion represents the adnominal region.}
  \label{fig:bodyshape}
\end{figure}

The practical motivation of this paper is 
the assessment of human visceral adipose tissue (VAT) using body shape descriptors \citep[e.g.,][]{SamoDC13, SunXL17,  WangLZ19}. 
VAT is a type of body fat that is stored within the abdominal cavity and located near several vital organs such as liver, stomach, and intestines. Accurate quantitative assessment of VAT is clinically essential since it is known to be associated with the risk for several medial problems, 
including heart disease, Alzheimer's disease, type 2 diabetes, stroke, insulin resistance and high cholesterol among others \citep[e.g.,][]{Desp07, JungHK16}. Commonly used medical equipments which can provide accurate VAT measurements include 
dualenergy X-ray absorptiometry (DXA), computed tomography (CT) and magnetic resonance imaging. However, using them is always expensive and reliant on professional operations, and some equipments, e.g., DXA and CT, may expose patients to ionizing radiation \citep[e.g.,][]{YehSW09}. 
Due to the prevalence of commodity optical body scan systems, body shape provides a safe and affordable alternative to assessing VAT volume
\citep[e.g.,][]{NgHF16}. 
Figure \ref{fig:bodyshape} provides an illustration of body circumference from neck and ankle, a body shape descriptor extracted from 3D body geometry captured by a commodity level optical body scanner 
\citep{LuHZ19, WangLZ19}. Apparently body circumference may be regarded as a functional variable of which input is each level between neck and ankle. 
Following the framework by \citet{Neym23} and \citet{Rubi74}, body shape may be regarded as a treatment since people are able to physically intervene it, or parts of it,  
e.g., via dietary management and exercise, but researchers have so far only focused on its correlation with VAT rather than their causal relationship. 
In this paper we will fill this void and apply our proposed method to study the causal effect of body shape on VAT.

The main contribution of this paper is twofold. First, the propensity score for functional treatments is properly defined, termed ``functional propensity score'', to balance covariates. When a treatment is binary, multi-level categorical or continuous, the propensity score \citep{RoseR83} or generalized propensity score \citep{Imbe00, HiraI04, ImaivD04} has been commonly used to balance covariates, which is defined in terms of the conditional mass/density function of the treatment given covariates. However, this way of defining propensity scores is inapplicable to functional treatments since the density function for a functional variable generally does not exist \citep{DelaH10}. To circumvent this problem, we propose to substitute the functional treatment by its top functional principal component (FPC) scores, a multivariate variable, and define 
the functional propensity score by the conditional density function of the FPC scores given covariates. Second, we propose two covariate balancing methods to estimate the functional propensity score. Traditionally the propensity score is estimated by a parametric model together with maximum likelihood, but it is well known that this approach may suffer from model misspecification substantially \citep[e.g.,][]{KangS07}. 
To alleviate this drawback, covariate balancing methods with the aim to mimic randomization have been very popular recently \citep[e.g.,][]{QinZ07, Hain12, ImaiR14, 
Zubi15, ChanYZ16, 
LiMZ18, 
WongC18, YiuS18, 
Zhao19}. 
Following the same idea, we seek functional propensity score estimates which can 
minimize the correlation between the FPC scores and covariates. Our methods generalize the parametric and nonparametric covariate balancing methods by \citet{FongHI18} to estimate the 
generalized propensity score for continuous treatments, which can improve likelihood-based parametric estimation and avoid parametric misspecification respectively. 
With a functional propensity score estimate, we fit the outcome model by inverse probability weighting \citep{Rose87, RobiHB00, HiraIR03}, which is closely related to the covariate balancing measure.

The rest of the paper proceeds as follows. Section \ref{sec:fps} provides the definition of the functional propensity score. In Section \ref{sec:method}, two covariate balancing methods are proposed to estimate the functional propensity score, and the causal effect estimation by functional propensity score weighting is also introduced. Section \ref{sec:simu} presents a simulation study to evaluate the performances of the two proposed methods with respect to covariate balancing and causal effect estimation accuracy. In Section \ref{sec:data}, the proposed methods are applied to study the causal effect of body circumference on VAT. Section \ref{sec:dis} concludes the paper.

\section{Functional Propensity Score}\label{sec:fps}

Let $Y$ be a continuous outcome, the treatment $X$ be a smooth, e.g., twice-differentiable, random function defined on a compact domain $\mT$ which is square-integrable, i.e., $E \{ \int_{\mT} X^2(t)\, dt \} < \infty$, and $\bC$ be a $p-$dimensional multivariate covariate with $E (\bC^\top \bC) < \infty$. Without loss of generality we assume $\mT=[0, 1]$, $E\{X(t)\}=0, t \in [0, 1]$ and $E(\bC) = \bzero$. Suppose that we observe $n$ samples $\{(Y_i, X_i, \bC_i): i=1, \ldots, n\}$ in practice, which are independently and identically distributed (i.i.d.) copies of $(Y, X, \bC)$. 
For simplicity the trajectories of all $\{X_i: i=1, \ldots, n\}$ are assumed fully observable and uncontaminated by noise. 
We aim to estimate $E\{Y(x)\}$ in terms of $\{(Y_i, X_i, \bC_i): i=1, \ldots, n\}$, where $x \in L^2=\{f: \int_0^1 f^2(t)\, dt < \infty\}$ is any functional value the treatment $X$ can take.

Following the classical causal inference literature, we assume
strong ignorablity. 
\begin{asmp}[Strong Ignorability] \label{asmp:ignore}
Let $Y(x)$ be the potential outcome given the treatment value $X=x$. 
Assume
$
X \perp Y(x) \mid \bC$ for all $x \in L^2$,
where 
``$\perp$'' represents independence.
\end{asmp}

The propensity score 
or generalized propensity score for a categorical or continuous treatment is defined in terms of the conditional probability mass/density function of the treatment given covariates. 
However, this way of defining the propensity score 
is inapplicable for functional treatments since the probability density function for a functional variable generally does not exist \citep{DelaH10}. 
Therefore we 
propose to define the propensity score in terms of functional principal component (FPC) scores of the functional treatment. 

Explicitly, the Karhunen-Lo\`{e}ve theorem allows for the representation  
$
X(t) = \sum_{k=1}^\infty A_k \phi_k(t)$, $t \in [0, 1], 
$
where $\phi_k, k \geq 1$ are eigenfunctions corresponding to the eigenvalues $\lambda_1 \geq \lambda_2 \geq \cdots > 0$ of the covariance function $\text{Cov}\{X(s), X(t)\}$. The FPC scores $A_k=\int_0^1 X(t) \phi_k(t), k \geq 1$ are mutually uncorrelated with zero means and $\text{Var}(A_k) = \lambda_k, k \geq 1$. 
Typically $X$ can be well approximated by the first $L$ summands, i.e., 
$
X(t) \approx \sum_{k=1}^L A_k \phi_k(t),
$
where $L$ is often determined such that the top $A_k, k =1, \ldots, L$ 
cumulatively account for a large proportion of variation of $X$, e.g., 95\% or 99\%. Since the majority of $X_i$'s information can be captured by $\bA=(A_1, \ldots, A_L)^\top$, 
we may use $\bA$, a multivariate substitute for $X$, to define the propensity score. Apparently if $X$ is of finite rank $L$, then $\bA$ 
can fully capture the information on $X$.



We next define the propensity score for functional treatments in terms of $\bA$, an $L$-dimensional substitute of $X$.

\begin{defi}[Functional Propensity Score]\label{defi:FPS}

Let $r$ be the conditional probability density function of $\bA$ given $\bC$, i.e., 
$
r(\bba, \bc) = f_{\bA \mid \bC} (\bba \mid \bc).
$
The rank-$L$ functional propensity score (FPS) 
is 
defined by $e_i=r(\bA_i, \bC_i)$, 
$, i=1, \ldots, n$. 
The corresponding rank-$L$ FPS weight is defined by 
$w_i = f_{\bA}(\bA_i)/ e_i, i=1, \ldots, n$, 
where $f_{\bA}(\cdot)$ is the marginal density of $\bA$.
\end{defi}

In practice, one may standardize data before the analysis. Since $A_1, \ldots, A_L$ are mutually uncorrelated with $\text{Var}(A_k)=\lambda_k, k=1, \ldots, L$, the standardized $\bA$ is $\bA^* = (A_1^*, \ldots, A_L^*)^\top$ 
where $A_k^*=\lambda_k^{-1/2} A_k, k=1, \ldots, L$. Let $\bC^*=\Gamm_\bC^{-1/2}\bC$ be the standardized $\bC$ where $\Gamm_\bC=E(\bC \bC^\top)$. 
Similarly for the $i$th subject, denote $\bC_i^*=\Gamm_\bC^{-1/2}\bC_i$ and $\bA_i^* = (A_{i1}^*, \ldots, A_{iL}^*)^\top$ where 
$A_{ik}=\lambda_k^{-1/2} \int_0^1 X_i(t) \phi_k(t), k=1, \ldots, L$. 
Then the functional propensity score 
can be alternatively defined in terms of $\bA^*$ and $\bC^*$.


\begin{defi}[Standardized Functional Propensity Score]\label{defi:SFPS}
Let $r^*$ be the conditional probability density function of $\bA^*$ given $\bC^*$, i.e., 
$
r^*(\bba, \bc) = f_{\bA^* \mid \bC^*} (\bba \mid \bc).
$
The rank-$L$ standardized functional propensity score (SFPS) 
is defined by $e_i^*=r^*(\bA_i^*, \bC_i^*)$, 
$, i=1, \ldots, n$. 
The corresponding rank-$L$ SFPS weight is 
$w_i^* = f_{\bA^*}(\bA_i^*)/ e_i^*, i=1, \ldots, n$, 
where $f_{\bA^*}(\cdot)$ is the marginal density of $\bA^*$.
\end{defi}

We further assume positivity for both FPS and SFPS.

\begin{asmp}[Positivity]\label{asmp:posi}
For $r$ and $r^*$ in Definitions \ref{defi:FPS} and \ref{defi:SFPS} respectively, assume that $r(\bba, \bC_i),  r^*(\bba, \bC_i^*) > 0$ for all $\bba \in \mbR^L$.
\end{asmp}

For simplicity, hereafter we will only use SFPS to remove or mitigate the imbalance between the functional treatment and covariates. 

\begin{rmk} \label{rmk:FPS}
\item 1. Note that $\phi_k, \lambda_k, A_{ik}, k \geq 1, i=1, \ldots, n$ 
and $\Gamm_\bC$ above are population quantities. In practice we can only obtain their sample counterparts, but for brevity we abuse these notations to denote either the population or sample versions depending on their corresponding contexts.
\item 2. In practice functional data are always discretely measured and may be contaminated by noise. For densely measured functional data, one may 
pre-smooth each function to accurately recover its trajectory \citep[e.g.,][]{ZhanC07} and the FPC scores 
accordingly. 
\item 3. To achieve a satisfactory balance between the functional treatment and covariates, the rank $L$ of both FPS and SFPS is supposed to be sufficiently large such that a majority of the variation of the functional treatment is represented by its top FPC scores.
The most commonly used method for selecting $L$ is via the percentage of variance explained by the top FPC scores, which is typically set as 95\% or 99\%. Note that the selection of $L$ is constrained by the sample size and number of covariates, 
which will be addressed in Remark \ref{rmk:method}. 
\end{rmk}

\section{Methodology}\label{sec:method}

In this section we propose two covariate balancing methods to estimate the rank-$L$ FPS weights and 
the corresponding causal effect estimation by FPS weighting. 

\subsection{SFPS Estimation} \label{sec:sfpsest}
Without loss of generality we focus on $\bA_i^*$ and $\bC_i^*$, $i=1, \ldots, n$ after standardization. 
To balance covariates, we seek weights to minimize the weighted correlation between $\bA_i^*$and $\bC_i^*$. By easy calculations we can obtain 
\begin{equation}\label{eq:cb}
E w_i^* \bA_i^* (\bC_i^*)^\top = \bzero, \quad i=1, \ldots, n.
\end{equation}
Thus  
the rank-$L$ SFPS weight $w_i^* = f_{\bA^*}(\bA_i^*)/ f_{\bA^* \mid \bC^*}(\bA_i^* \mid \bC_i^*), i=1, \ldots, n$ is a minimizer.

Motivated by \citet{FongHI18} who focused on continuous treatments, we consider a parametric and a nonparametric method to estimate rank-$L$ SFPS weights with the covariate balancing condition (\ref{eq:cb}) taken into account.


\noindent 
\textbf{Parametric SFPS Estimation.} 
To estimate 
$w_i^* = f_{\bA^*}(\bA_i^*)/ f_{\bA^* \mid \bC^*}(\bA_i^* \mid \bC_i^*), i=1, \ldots, n$, we assume parametric structures on $f_{\bA^*}$ and $f_{\bA^* \mid \bC^*}$ and estimate involved unknown parameters. 
First we assume that $\bA^*$ is jointly normal. Then since the components of $\bA^*$, which are standardized FPC scores, are mutually uncorrelated, 
$f_{\bA^*}$ is known as 
$$
f_{\bA^*}(\bA_i^*) = (2 \pi)^{-L/2} \exp\left\{ -\frac{1}{2}  \left(\bA_i^* \right)^\top \left(\bA_i^* \right)\right\}.
$$
We further assume that $\bA^*$ given $\bC^*$ is also jointly normal. Then there exist unknown matrices $\betam$ and $\Sigmam$ such that 
$$
f_{\bA^* \mid \bC^*}(\bA_i^* \mid \bC_i^*) = (2 \pi)^{-L/2} \det\left( \Sigmam\right)^{-1/2} \exp\left\{ -\frac{1}{2}  \left(\bA_i^* - \betam^\top \bC_i^* \right)^\top \Sigmam^{-1}  \left(\bA_i^* - \betam^\top \bC_i^* \right)\right\}, 
$$
\begin{equation}\label{eq:sfpspara}
\text{so}\quad w_i^* = 
\det\left( \Sigmam\right)^{1/2} \exp\left\{ \frac{1}{2}  \left(\bA_i^* - \betam^\top \bC_i^* \right)^\top \Sigmam^{-1}  \left(\bA_i^* - \betam^\top \bC_i^* \right) - \frac{1}{2}  \left(\bA_i^* \right)^\top \left(\bA_i^* \right)\right\}.
\end{equation}


Based on the fact $E\left(\bA_i^* - \betam^\top \bC_i^* \right)\left(\bA_i^* - \betam^\top \bC_i^* \right)^\top= \Sigmam$ together with (\ref{eq:cb}), 
we solve the following method-of-moments equations to estimate the unknown parameters $\betam$ and $\Sigmam$:
\begin{equation*}
\begin{cases} 
n^{-1}\sum_{i=1}^n \left(\bA_i^* - \betam^\top \bC_i^* \right)\left(\bA_i^* - \betam^\top \bC_i^* \right)^\top= \Sigmam; \\ 
n^{-1}\sum_{i=1}^n  \det\left( \Sigmam\right)^{1/2} \exp\left\{ \frac{1}{2}  \left(\bA_i^* - \betam^\top \bC_i^* \right)^\top \Sigmam^{-1}  \left(\bA_i^* - \betam^\top \bC_i^* \right) - \frac{1}{2}  \left(\bA_i^* \right)^\top \left(\bA_i^* \right)\right\} \bA_i^* (\bC_i^*)^\top = \bzero.
\end{cases}
\end{equation*}
With the estimates of $\betam$ and $\Sigmam$, 
$w_i^*, i=1, \ldots, n$ can be 
estimated following (\ref{eq:sfpspara}).
\noindent 
\textbf{{Nonparametric SFPS Estimation}.} 
To avoid possible parametric misspecification for the SFPS, one may adopt the empirical likelihood method \citep{Owen01} instead where no parametric assumption is needed.  
Simple calculations can show that 
$w_i^* = f_{\bA^*}(\bA_i^*)/ f_{\bA^* \mid \bC^*}(\bA_i^* \mid \bC_i^*), i=1, \ldots, n$ satisfies the four conditions below including (\ref{eq:cb}): 
\begin{equation}\label{eq:cond}
E\left( w_i^*\right) = 1, \  E\left\{ w_i^* \bA_i^* (\bC_i^*)^\top\right\} = \bzero, \ E\left( w_i^*\bA_i^* \right) = \bzero, \ \text{and} \  E\left( w_i^* \bC_i^*\right) = \bzero, \ i=1, \ldots, n.
\end{equation}
Subject to the empirical counterparts of (\ref{eq:cond}), we aim to maximize 
$
\prod_{i=1}^n f_{(\bA^*, \bC^*)}(\bA_i^*, \bC_i^*), 
$
where $f_{(\bA^*, \bC^*)}$ is the joint density of $\bA^*$ and $\bC^*$. Since $f_{(\bA^*, \bC^*)}(\bA_i^*, \bC_i^*)=\{f_{\bA^*}(\bA_i^*)f_{\bC^*}(\bC_i^*)\}/w_i^*$, an equivalent optimization is
\begin{equation} \label{eq:npexact}
\min_{\bw^*} \sum_{i=1}^n \log(w_i^*),\quad
\text{s.t.}  \sum_{i=1}^n w_i^* =n,\  \sum_{i=1}^n w_i^* \bA_i^* (\bC_i^*)^\top =\bzero, \  \sum_{i=1}^n w_i^* \bA_i^* =\bzero, \   \sum_{i=1}^n w_i^* \bC_i^* = \bzero,
\end{equation}
where $\bw^*=(w_1^*,\ldots, w_n^*)^\top$. 
By the method of Lagrangian multipliers, it is easy to show that the constrained optimization (\ref{eq:npexact}) 
is equivalent to the unconstrained optimization
\begin{equation*}
\max_{\gammam} \sum_{i=1}^n \log \left(1-\gammam^\top \bg_i \right), \quad \text{where $\bg_i = ((\bA_i^*)^\top, (\bC_i^*)^\top, \text{vec}(\bA_i^* (\bC_i^*)^\top)^\top)^\top$},
\end{equation*}
with ``$\text{vec}$'' referring to the vectorization operation, 
which is unfortunately non-convex. 

To resolve this non-convexity issue, we adopt the regularized approach by \citet{FongHI18} which 
allows for an imbalance
between $\bA_i^*$ and $\bC_i^*$ but meanwhile penalizes such imbalance in the objective function. 
Explicitly, we consider an $\ell^2$-regularized optimization
\begin{eqnarray}\label{eq:npreg} 
&& \min_{\bw^*, \Gamm} \left[\sum_{i=1}^n \log(w_i^*) + \frac{1}{2 \rho} \{\text{vec}(\Gamm)\}^\top \{\text{vec}(\Gamm)\} \right], \\
\text{s.t.} &&  \sum_{i=1}^n w_i^* =n,\  \frac{1}{n}\sum_{i=1}^n w_i^* \bA_i^* (\bC_i^*)^\top =\Gamm, \  \sum_{i=1}^n w_i^* \bA_i^* =\bzero,  \  \sum_{i=1}^n w_i^* \bC_i^* = \bzero, \nonumber
\end{eqnarray}
where $\rho > 0$ is a tuning parameter. Obviously the constraint $n^{-1}\sum_{i=1}^n w_i^* \bA_i^* (\bC_i^*)^\top =\Gamm$ in (\ref{eq:npreg}) relaxes the sample counterpart of (\ref{eq:cb}), which allows for an imperfect balance between $\bA_i^*$ and  $\bC_i^*$. Meanwhile, such imbalance is regularized via an $\ell^2$-penalty and a smaller tuning parameter $\rho$ tends to shrink $\Gamm$ towards zero.


Note that $\Gamm$ is a $L \times p$ matrix and the optimization (\ref{eq:npreg}) is difficult due to its high dimensionality. To reduce its dimensionality, instead of searching all possible values of $\Gamm$, we let $\Gamm = \theta \Gamm_0$ where $\theta \in [-1, 1]$ is a scalar to be optimized and $\Gamm_0 = n^{-1}\sum_{i=1}^n \bA_i^* (\bC_i^*)^\top$ is the unweighted covariance matrix between $\bA_i^*$ and $\bC_i^*$. In other words we only consider those weights which can uniformly reduce the entrywise imbalance between $\bA_i^*$ and $\bC_i^*$ to a certain degree. Accordingly the optimization (\ref{eq:npreg}) becomes 
\begin{eqnarray}\label{eq:npregsimp}
&& \min_{\bw^*, -1 \leq \theta \leq 1} \left[ \sum_{i=1}^n \log(w_i^*) + \frac{\theta^2}{2 \rho} \{\text{vec}(\Gamm_0)\}^\top \{\text{vec}(\Gamm_0)\} \right], \\
\text{s.t.} &&  \sum_{i=1}^n w_i^* =n,\  \frac{1}{n}\sum_{i=1}^n w_i^* \bA_i^* (\bC_i^*)^\top = \theta \Gamm_0, \  \sum_{i=1}^n w_i^* \bA_i^* =\bzero,  \  \sum_{i=1}^n w_i^* \bC_i^* = \bzero, \nonumber
\end{eqnarray}
By the Lagrangian multiplier and profile method, the optimization (\ref{eq:npregsimp}) can be solved by a double-loop routine: 
\begin{eqnarray*}
\text{Inner Loop}&:& \hgammam(\theta) = \argmax_{\gammam} \sum_{i=1}^n \log \left\{ 1-\gammam^\top \bh_i(\theta)\right\},\quad \text{for any fixed $\theta \in [-1, 1]$}; \\
\text{Outer Loop}&:& \hat{\theta} = \argmax_{-1 \leq \theta \leq 1} \left[ \sum_{i=1}^n \log \left\{ 1-\hgammam(\theta)^\top \bh_i(\theta)\right\}- \frac{\theta^2}{2 \rho} \{\text{vec}(\Gamm_0)\}^\top \{\text{vec}(\Gamm_0)\} \right],\\
&&\text{and} \quad \hw_i^* = \frac{1}{1-\hgammam(\hat{\theta})^\top \bh_i(\hat{\theta})}, \quad i=1,\ldots, n,
\end{eqnarray*}
where $\bh_i(\theta)=((\bA_i^*)^\top, (\bC_i^*)^\top, \text{vec}(\bA_i^* (\bC_i^*)\top - n \theta \Gamm_0)^\top)^\top, i=1, \ldots, n$.
The inner loop can be solved by the Broyden--Fletcher--Goldfarb--Shanno algorithm while $\hat{\theta}$ in the outer loop can be found by a grid search.

\begin{rmk}\label{rmk:method}
\item 1. 
Both parametric and nonparametric SFPS estimation methods above generalize the 
methods by \citet{FongHI18} for continuous treatments since a functional treatment is degenerated into a continuous treatment if $\bA^*$ is one-dimensional, 
i.e., $L=1$.
 

\item 2. As in Remark \ref{rmk:FPS}, 
a large $L$ is preferred to achieve a satisfactory covariate balance, but its practical selection is constrained by the sample size $n$ and the number of covariates $p$. The parametric method involves estimating $\betam$ and $\Sigmam$, of which dimensions are $p \times L$ and $L \times L$ respectively. The nonparametric method involves balancing the covariance matrix between $\bA^*$ and $\bC^*$, of which dimension is $L \times p$. Therefore both estimations will be computationally difficult if a too large $L$ is selected. 

\item 3. In the nonparametric SFPS estimation, solving (\ref{eq:npregsimp}) involves tuning $\rho$ properly. Generally a suitable $\rho$ should be small since otherwise a large $\htheta$ is typically obtained, which leads to a liberal imbalance. However, if $\rho$ is extremely small to indicate a very low tolerance of imbalance between $\bA^*$ and $\bC^*$, algorithmic convergence is likely to fail, which also leads to a poor covariate balance. 
Following 
\citet{FongHI18}, 
we set a default value $\rho = 0.1/n$, but a practitioner may need to explore multiple values to achieve a satisfactory balance.
\end{rmk}


\subsection{Causal Effect Estimation} \label{sec:ce}

To estimate the causal effect of the functional treatment on the scalar outcome, 
one may fit a scalar-on-function outcome model via propensity score weighting. 
The SFPS weight estimates obtained by either the parametric or nonparametric method in Section \ref{sec:sfpsest} are used as weights in the involved objective function such that the bias due to the imbalance between the functional treatment and covariates can be substantially reduced.


For example, if the outcome model is assumed a functional linear model 
\begin{equation} \label{eq:mFLRf}
E(Y \mid X) = \mu_0 + \int_0^1 \mu(t) X(t)\, dt,
\end{equation}
with the smooth causal effect $\{\mu(t): t \in [0, 1]\}$ and $\mu_0 = E(Y)$ since $E\{X(t)\}=0$ for all $t \in [0, 1]$, 
one may estimate $\mu$ is via the following truncation method. 
Let $\{\psi_k(t): t\in [0, 1]; k=1, \ldots, L^*\}$ be a set of smooth and orthonormal $L^2$ basis functions such that $\mu(t) \approx \sum_{k=1}^{L^*} \mu_k \psi_k(t), t \in [0, 1]$ where $\mu_k = \int_0^1 \mu(t) \psi_k(t)\, dt, k=1, \ldots, L^*$. 
Then 
$
E(Y \mid X) \approx \mu_0 + \sum_{k=1}^{L^*}  \mu_k \{ \int_0^1 X(t) \psi_k(t)\, dt \} 
$
and the causal effect can be estimated by $\hmu(t) = \sum_{k=1}^{L^*}  \hmu_k \psi_k(t), t \in [0, 1]$, where
\begin{equation} \label{eq:wls}
(\hmu_1,\ldots, \hmu_{L^*}) = \argmin_{\mu_1, \ldots, \mu_{L^*}} \sum_{i=1}^n \hw_i \left(Y_i - \bar{Y} - \sum_{k=1}^{L^*}  \mu_k B_{ik}\right)^2,
\end{equation}
with $\bar{Y} = n^{-1}\sum_{i=1}^n Y_i$, $B_{ik}= \int_0^1 X_i(t) \psi_k(t)\, dt$, $k=1, \ldots, L^*$, $i=1, \ldots, n$, and $\hw_i, i=1, \ldots, n$ being estimated SFPS weights obtained by either the parametric or nonparametric method in Section \ref{sec:sfpsest}. 




It is very common that the basis functions $\{\psi_k: k=1, \ldots, L^*\}$ are chosen as the eigenfunctions of $X$ and accordingly 
$B_{ik}$ $k=1, \ldots, L^*$, $i=1, \ldots, n$ are FPC scores. 
Note that the $L^*$ FPC scores used to fit the outcome model
may be chosen independently and differently from those $L$ FPC scores used to define FPS for covariate balancing. Different from $L$ of which selection does not involve the outcome, 
the $L^*$ FPC scores used to fit an outcome model may be selected in terms of their predictability of the outcome. We will illustrate this numerically in Section \ref{sec:simu} below.

\section{Simulation} \label{sec:simu}

%
%

In this section we present a simulation study to evaluate the numerical performance of the proposed method. We had $200$ simulation runs with $n=200$ independent subjects in each run. The noiseless functional treatment for the $i$th subject was generated by $X_i(t) = \sum_{k=1}^6 A_{ik}\phi_k(t), t \in [0,1]$ where the eigenfunctions are $\phi_{2k-1}(t) = \sqrt{2} \textrm{sin}(2\pi kt)$ and $\phi_{2k}(t) = \sqrt{2} \textrm{cos}(2\pi kt)$ for $k = 1, 2, 3$, and the FPC scores are $A_{i1} = 4Z_{i1}$, $A_{i2} = 2\sqrt{3}Z_{i2}$, $A_{i3} = 2\sqrt{2}Z_{i3}$, $A_{i4} = 2Z_{i4}$, $A_{i5} = Z_{i5}$, and 
$A_{i6} = Z_{i6} / \sqrt{2}$, with $Z_{i1}, \ldots, Z_{i6}$ independently sampled from the standard normal distribution.

We considered the following four settings to generate the outcome $Y_i$ and a three-dimensional covariate $\bC_i = (C_{i1}, C_{i2}, C_{i3})^\top$ depending on whether the FPC scores or outcome are linear in the covariate. 

\begin{itemize}
\item Setting 1. The covariate 
was generated by $C_{i1} = Z_{i1}+W_{i1}$, $C_{i2} = 0.2 Z_{i2} + W_{i2}$, and $C_{i3} = 0.2Z_{i3} + W_{i3}$ where $W_{i1} \sim N(0, 1)$, $W_{i2} \sim N(0, 0.5)$ and $W_{i3} \sim N(0, 0.5)$ are independent. 
The outcome was obtained by
$$
Y_i = 1 + \int_0^1 \mu(t)X_i(t)\, dt + 2C_{i1} + e_i, 
$$ 
with the true causal effect 
$$
\mu(t) = 2\sqrt{2}\textrm{sin}(2\pi t) + \sqrt{2} \textrm{cos}(2\pi t) + \sqrt{2} \textrm{sin}(4\pi t)/2 + \sqrt{2}\textrm{cos}(4\pi t)/2, \quad t \in [0,1], 
$$ 
and the noise $e_i \sim N(0, 25)$. In this setting, both the FPC scores $(A_{i1}, \ldots, A_{i6})^\top$ and outcome $Y_i$ are linear in $\bC_i$. 

\item Setting 2. The covariate was generated by $C_{i1} = (Z_{i1}+0.5)^2+W_{i1}$, $C_{i2} = 0.2 Z_{i2} + W_{i2}$, and $C_{i3} = 0.2Z_{i3} + W_{i3}$ where $W_{i1} \sim N(0, 1)$, $W_{i2} \sim N(0, 0.5)$ and $W_{i3} \sim N(0, 0.5)$ are independent, so the FPC scores are nonlinear in the covariate. The outcome was generated the same as in Setting 1. 

\item Setting 3. The covariate was generated in Setting 1. The outcome was obtained by
$$
Y_i = 1 + \int_0^1 \mu(t)X_i(t)\, dt + 2C_{i1} + C_{i2}^2 + e_i,
$$
where $e_i \sim N(0, 25)$, so the outcome is nonlinear in the covariate.


\item Setting 4. The covariate was generated in Setting 2 while the outcome was generated the same as in Setting 3, so both the FPC scores and outcome are nonlinear in the covariate.
\end{itemize}

\begin{figure}[H]
    \centering
    \includegraphics[width=12cm]{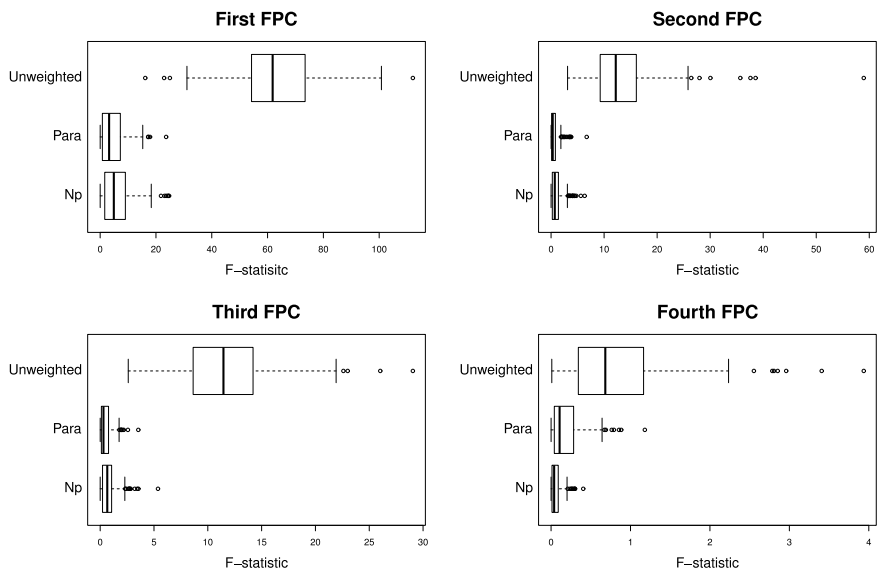}
    \caption{Boxplots of F-statistics for linear regression models of selected top FPC scores on the covariate for Settings 1 and 3. The top FPC scores were selected by $\text{PVE}_L = 0.95$. Each linear regression model was fitted by ordinary least squares (Unweighted) or weighted least squares where weights were the estimated SFPS weights obtained by either the parametric (Para) or nonparametric (Np) covariate balancing method. 
    }
    \label{fig:ss3_bal95}
\end{figure}

\begin{figure}[H]
    \centering
    \includegraphics[width=12cm]{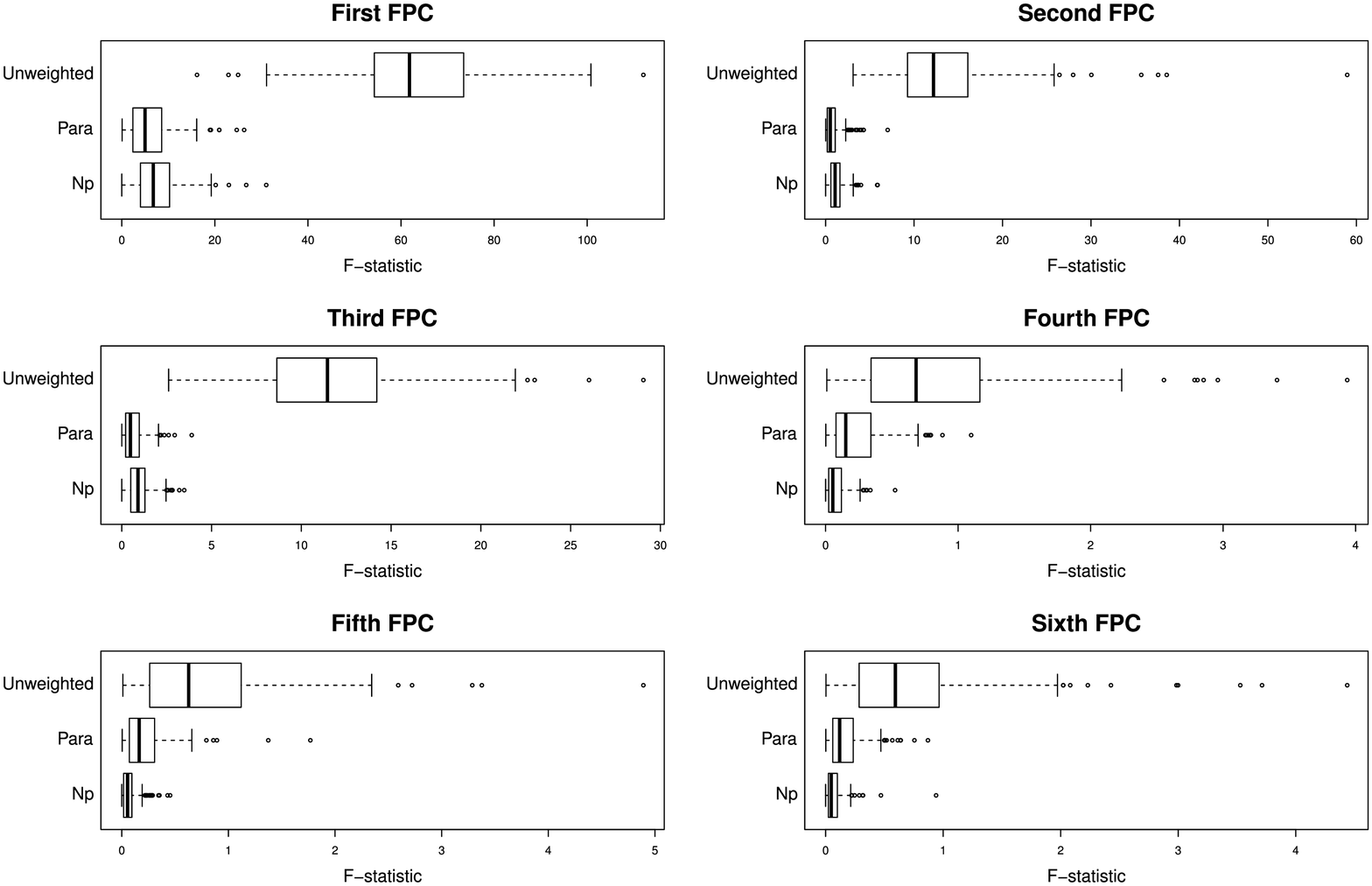}
    \caption{The same as Figure \ref{fig:ss3_bal95} except for $\text{PVE}_L = 0.99$. The boxplots for the sixth FPC score were created based on 
    187 simulation runs while those for the other five FPC scores were based on all $200$ simulation runs. }
    \label{fig:ss3_bal99}
\end{figure}

\begin{figure}[H]
    \centering
    \includegraphics[width=12cm]{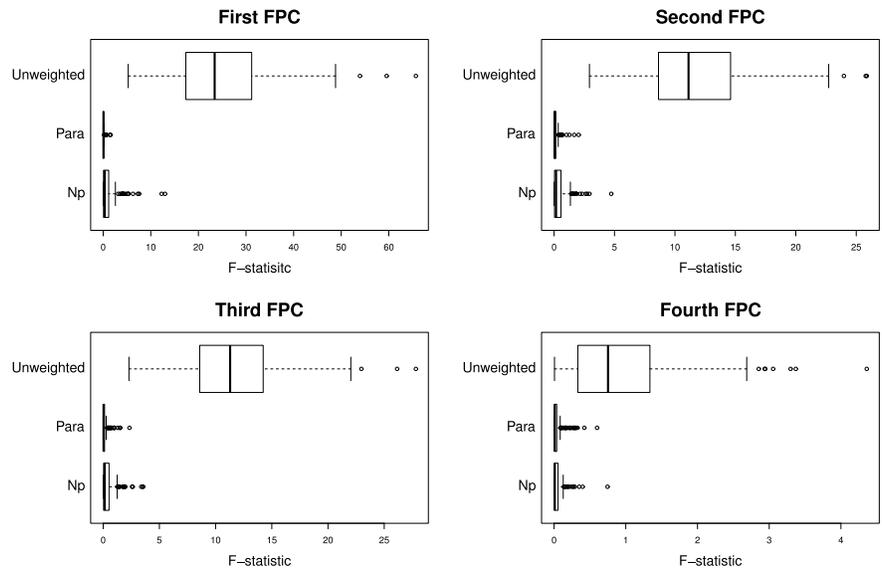}
    \caption{The same as Figure \ref{fig:ss3_bal95} except for Settings 2 and 4.
    } 
    \label{fig:ss4_bal95}
\end{figure}

\begin{figure}[H]
    \centering
    \includegraphics[width=12cm]{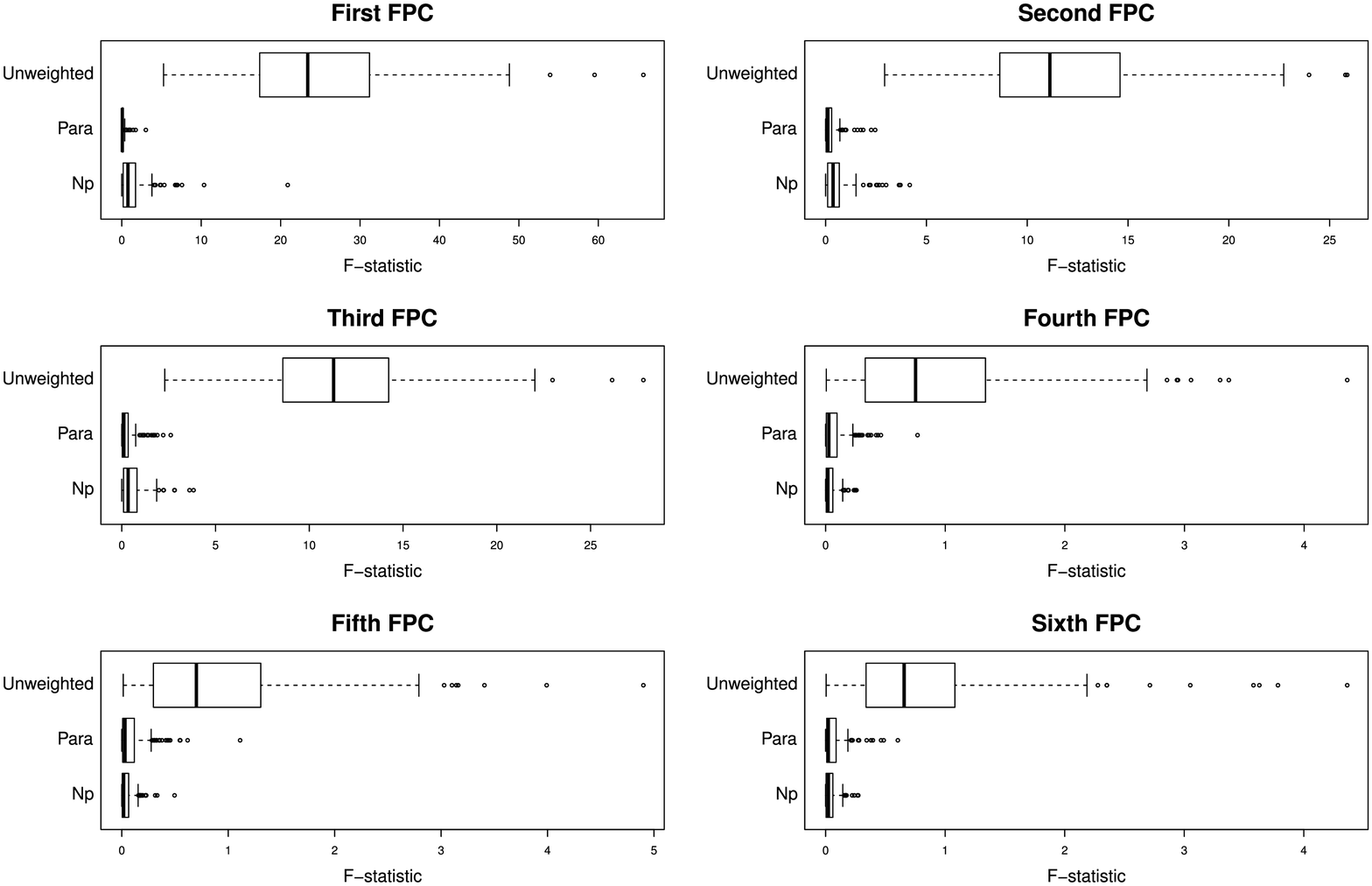}
    \caption{The same as Figure \ref{fig:ss4_bal95} except for $\text{PVE}_L = 0.99$. The boxplots for the sixth FPC score were created based on 
    187 simulation runs while those for the other five FPC scores were based on all $200$ simulation runs.}
    \label{fig:ss4_bal99}
\end{figure}

In each setting, the percentage of variance explained (PVE) was set as $\text{PVE}_L=0.95$ and $0.99$ respectively to 
select $L$ top FPC scores of $X$ to define the rank-$L$ FPS and SFPS. The corresponding SFPS weights were estimated by 
both the parametric and nonparametric covariate balancing methods in Section \ref{sec:sfpsest} respectively. The quality of imbalance reduction 
between the FPC scores and covariate is evaluated by the F-statistic for the linear regression of each selected top FPC score on the covariate, which was fitted by weighted least squares with the estimated SFPS weights obtained by the parametric or nonparametric method. The F-statistic for each linear regression fitted by ordinary least squares is also given for comparison.
The boxplots of these F-statistics are given in Figures \ref{fig:ss3_bal95}, \ref{fig:ss3_bal99},  \ref{fig:ss4_bal95}, and \ref{fig:ss4_bal99}. 

All four figures demonstrate a substantial and appealing imbalance reduction between the functional treatment and covariate due to either parametric or nonparametric SFPS weighting for both $\text{PVE}_L = 0.95$ and $0.99$. The covariate balancing performances of the parametric and nonparametric methods are comparably satisfactory except in a few outlying simulation runs. The parametric method generally outperforms for the top three FPC scores while the nonparametric method is superior for the other FPC scores.


To estimate the causal effect of the functional treatment, the outcome model (\ref{eq:mFLRf}) was fitted 
by the truncation method in (\ref{eq:wls}) where the weights were the estimated SFPS weights obtained by the parametric or nonparametric method, the basis functions were chosen as the top $L^*$ eigenfunctions for $X$, and $L^*$ was determined by $\text{PVE}_{L^*}=0.95$ and $0.99$ respectively. To assess the accuracy for causal effect estimation, we calculated the integrated squared error (ISE) of each causal effect estimate in each simulation run and further the averaged ISE (AISE) over all simulation runs, together with the median ISE (MISE) due to the presence of outlying estimates. To evaluate the effect of bias removal due to covariate balancing, we computed the integrated squared bias (ISB) using causal effect estimates of all simulation runs. For comparison, all these summary statistics were also provided for 
causal effect estimates without adjustment between the functional treatment and covariate, each of which was obtained by fitting (\ref{eq:wls}) with all weights being one. 
These summary statistics for the four simulation settings are given in Tables \ref{tab:ss1}, \ref{tab:ss2}, \ref{tab:ss3}, and \ref{tab:ss4} respectively. 


\begin{table}[H]
\centering
\caption{Median ISE (MISE), averaged ISE (AISE), and integrated squared bias (ISB) values of causal effect estimates for Setting 1. Each causal effect estimate was obtained by fitting the outcome model (\ref{eq:mFLRf}) using the truncation method in (\ref{eq:wls}) where the weights were all one (Unweighted), the estimated SFPS weights obtained by either the parametric (Para) or nonparametric (Np) covariate balancing method. 
}
\label{tab:ss1}
\begin{tabular}{*{8}{c}}
  \multirow{2}*{} & \multirow{2}*{}   & \multicolumn{3}{c}{$\text{PVE}_{L^*} = 0.95$  } & \multicolumn{3}{c}{$\text{PVE}_{L^*} = 0.99$} \\
  \cmidrule(lr){3-5} \cmidrule(lr){6-8}
  
& &  MISE & AISE & ISB  & MISE & AISE & ISB  \\ 
  \hline
\multicolumn{2}{c}{Unweighted }& 0.3600 & 0.3767 & 0.2968 & 0.6397 & 0.7591 & 0.3042 \\
  \hline 
\multirow{2}*{$\text{PVE}_L = 0.95$ }  & Para & 0.2263 & 0.2553 & 0.0508 & 1.1476 & 1.6051 & 0.0772 \\ 
  & Np & 0.1938 & 0.2262 & 0.0539 & 0.8163 & 1.1478 & 0.0684 \\ 
  \hline
\multirow{2}*{$\text{PVE}_L = 0.99$}  &   Para & 0.2159 & 0.2419 & 0.0631 & 0.8002 & 1.1021 & 0.0987 \\ 
&  Np & 0.1943 & 0.2260 & 0.0712 & 0.5937 & 0.9010 & 0.0926 \\ 
   \hline
\end{tabular}
\end{table}


As shown in Tables \ref{tab:ss1}, \ref{tab:ss2}, \ref{tab:ss3}, and \ref{tab:ss4}, the ISB values for the causal effect estimators obtained by each covariate balancing method are always much smaller than those for the unadjusted one,  
which demonstrates the effectiveness of the proposed method to reduce the bias in causal effect estimation. 

\begin{table}[H]
\centering
\caption{The same as Table \ref{tab:ss1} except for Setting 2.}
\label{tab:ss2}
\begin{tabular}{*{8}{c}}
  \multirow{2}*{} & \multirow{2}*{}   & \multicolumn{3}{c}{$\text{PVE}_{L^*} = 0.95$ } & \multicolumn{3}{c}{$\text{PVE}_{L^*} = 0.99$} \\
  \cmidrule(lr){3-5} \cmidrule(lr){6-8}
  
& &  MISE & AISE & ISB  & MISE & AISE & ISB  \\ 
  \hline
\multicolumn{2}{c}{Unweighted }
& 0.3828 & 0.4143 & 0.2991 & 0.7782 & 0.9333 & 0.3068 \\ 
  \hline 
 
\multirow{2}*{$\text{PVE}_L = 0.95$}  & Para & 0.1421 & 0.1724 & 0.0055 & 1.0200 & 1.4399 & 0.0258 \\ 
  & Np & 0.1501 & 0.1885 & 0.0059 & 0.9622 & 1.4393 & 0.0122 \\ 
  \hline
  
\multirow{2}*{$\text{PVE}_L = 0.99$}  & Para & 0.1252 & 0.1549 & 0.0058 & 0.6571 & 0.8548 & 0.0440 \\ 
&  Np & 0.1444 & 0.1688 & 0.0158 & 0.5878 & 0.7998 & 0.0429 \\ 
   \hline
\end{tabular}
\end{table}

The MISE and AISE values in Tables \ref{tab:ss1}, \ref{tab:ss2}, \ref{tab:ss3}, and \ref{tab:ss4} 
show that the causal effect estimation accuracy is influenced by 
the choices of $\text{PVE}_L$ and $\text{PVE}_{L^*}$. First, for all three causal effect estimators, using $\text{PVE}_{L^*}=0.95$ to select top $L^*$ FPC scores in the outcome model always leads to a smaller MISE/AISE value and thus a better causal effect estimation than setting $\text{PVE}_{L^*}=0.99$ 
regardless of the choice of $\text{PVE}_L$. This is somewhat unsurprising since $\text{PVE}_{L^*}=0.95$ selects top $L^*=4$ FPC scores of $X$ in the outcome model (\ref{eq:mFLRf}), which is correctly specified as implied by the four simulation settings above. In this case, the causal effect estimation accuracies of the two covariate balanced estimators are comparable and substantially better than the unadjusted one. This observation is consistent with the discussion after (\ref{eq:wls}) in Section \ref{sec:ce} and highlights the importance of outcome model specification. 

\begin{table}[H]
\centering
\caption{The same as Table \ref{tab:ss1} except for Setting 3.}
\label{tab:ss3}
\begin{tabular}{*{8}{c}}
  \multirow{2}*{} & \multirow{2}*{}   & \multicolumn{3}{c}{$\text{PVE}_{L^*} = 0.95$ } & \multicolumn{3}{c}{$\text{PVE}_{L^*} = 0.99$} \\
  \cmidrule(lr){3-5} \cmidrule(lr){6-8}
  
& &  MISE & AISE & ISB  & MISE & AISE & ISB  \\ 
  \hline
\multicolumn{2}{c}{Unweighted }& 0.3596 & 0.3788 & 0.2982 & 0.6349 & 0.7617 & 0.3055 \\ 
  \hline 
 
\multirow{2}*{$\text{PVE}_L = 0.95$}  & Para & 0.2269 & 0.2564 & 0.0515 & 1.1306 & 1.6093 & 0.0772 \\  
  & Np & 0.1931 & 0.2281 & 0.0543 & 0.8099 & 1.1474 & 0.0689 \\ 
  \hline
  
\multirow{2}*{$\text{PVE}_L = 0.99$}  & Para & 0.2182 & 0.2437 & 0.0639 & 0.7905 & 1.1004 & 0.0988 \\  
&  Np & 0.2000 & 0.2277 & 0.0718 & 0.5952 & 0.9004 & 0.0934 \\ 
   \hline
\end{tabular}
\end{table}

Moreover, for the two adjusted causal effect estimators, selecting top $L$ FPC scores by a larger $\text{PVE}_L=0.99$ to define SFPS and balance covariates may improve the accuracy for causal effect estimation compared to $\text{PVE}_L=0.95$. 
The improvement is  substantial 
if the outcome model is misspecified, i.e., $\text{PVE}_{L^*}=0.99$. When $\text{PVE}_L=\text{PVE}_{L^*}=0.99$, in particular, 
the two adjusted causal effect estimators are more accurate than the unadjusted one in almost all settings. This phenomenon suggests that one use as many FPC scores as possible to define and estimate SFPS weights for the benefit of both reducing imbalance and enhancing causal effect estimation.

\begin{table}[H]
\centering
\caption{The same as Table \ref{tab:ss1} except for Setting 4.}
\label{tab:ss4}
\begin{tabular}{*{8}{c}}
  \multirow{2}*{} & \multirow{2}*{}   & \multicolumn{3}{c}{$\text{PVE}_{L^*} = 0.95$ } & \multicolumn{3}{c}{$\text{PVE}_{L^*} = 0.99$} \\
  \cmidrule(lr){3-5} \cmidrule(lr){6-8}
  
& &  MISE & AISE & ISB  & MISE & AISE & ISB  \\ 
  \hline
\multicolumn{2}{c}{Unweighted }
& 0.3844 & 0.4161 & 0.3005 & 0.7789 & 0.9331 & 0.3081 \\ 
  \hline 
 
\multirow{2}*{$\text{PVE}_L = 0.95$}  & Para & 0.1361 & 0.1754 & 0.0060 & 0.9833 & 1.4363 & 0.0261 \\   
  & Np & 0.1576 & 0.1903 & 0.0061 & 0.9856 & 1.4399 & 0.0127 \\
  \hline
  
\multirow{2}*{$\text{PVE}_L = 0.99$}  & Para & 0.1290 & 0.1572 & 0.0064 & 0.6338 & 0.8561 & 0.0436 \\   
&  Np & 0.1443 & 0.1719 & 0.0161 & 0.5685 & 0.8060 & 0.0437 \\  
   \hline
\end{tabular}
\end{table}

\section{Real Data Application} \label{sec:data}

We applied the proposed parametric and nonparametric covariate balancing methods to the VAT dataset introduced in Section \ref{sec:intro} where we aimed to estimate the causal effect of body shape on the VAT-to-weight ratio. The dataset consists of $90$ female and $60$ male subjects. For each subject, the VAT volume was obtained by a DXA scan and the body circumference sampled at $128$ equidistant levels from neck (level $1$) to ankle (level $128$) was extracted from an optical 3D body scan. 
See \citet{LuHZ19} for more data collection details. 
Since VAT is stored near organs in the abdominal region, we 
treated the circumference measured from level $30$ (below chest) to level $64$ (groin) as the functional treatment $X$, of which domain incorporates the abdominal region. 
The multivariate covariate includes ethnicity (white/non-white), the logarithm of height and its square. We centered all variables before the analysis.

\begin{figure}[H]
    \centering
    \includegraphics[width=12cm]{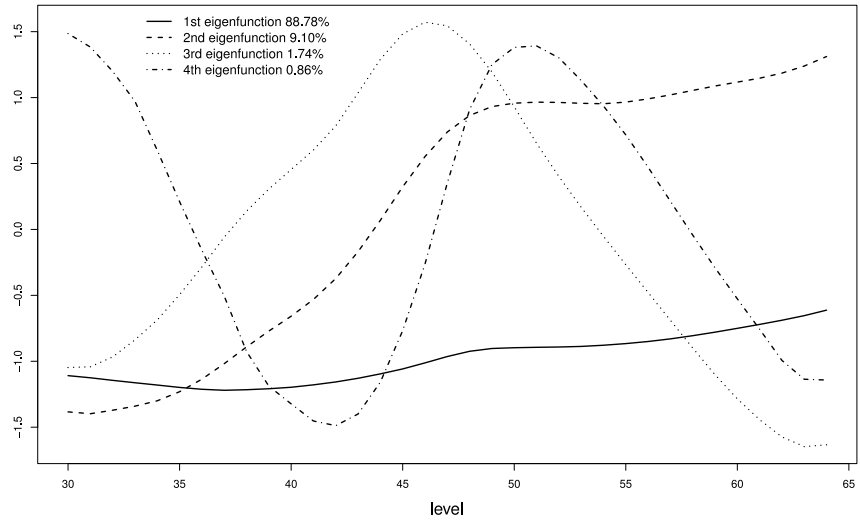}
    \caption{Top four eigenfunctions of the functional treatment and percentages of variation explained by their corresponding FPC scores. 
    } 
    \label{fig:efn}
\end{figure}

Following the suggestion in Section \ref{sec:simu} that one use FPC scores as many as possible to define SFPS weights, we selected the top $L=4$ FPC scores $\bA = \left( A_1, \ldots, A_L \right) ^\top$ of the functional treatment by setting $\text{PVE}_L=0.99$.
Their corresponding eigenfunctions are illustrated in Figure \ref{fig:efn}. Figure \ref{fig:efn} shows that the first eigenfunction plays a predominant role in the variation of $X$ with $87.88\%$ explained by the first FPC score, 
and that along the first eigenfunction, the variation in the upper part of the interested body region is larger than that in the lower part.

\begin{figure}[H]
    \centering
    \includegraphics[width=11cm]{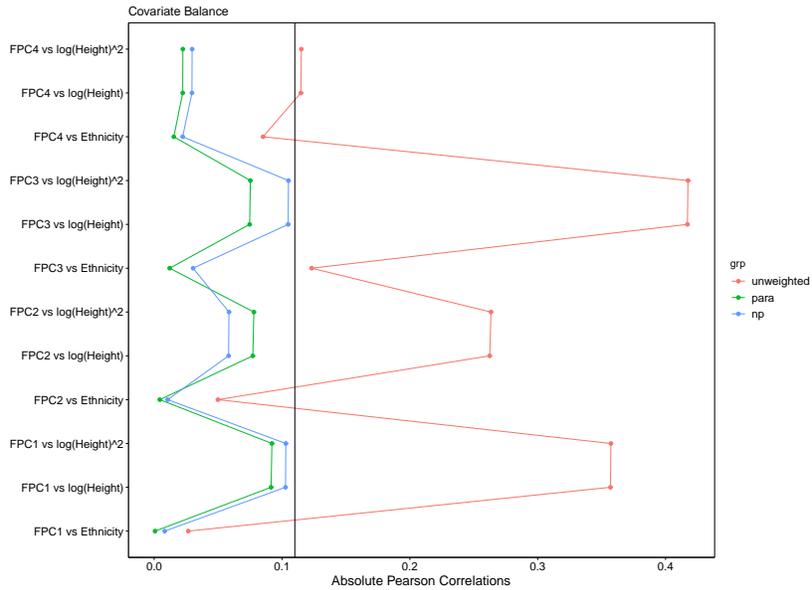}
    \caption{
    Absolute Pearson correlation between each FPC of the functional treatment and each covariate after weighting by either the parametric (Para) or nonparametric (Np) covariate balancing method. The unweighted absolute Pearson correlation (Unweighted) is also given for comparison. 
    } 
    \label{fig:data_bal}
\end{figure}

To assess the quality of the proposed method in improving covariate balance, we calculated the absolute weighted Pearson correlation between each of the four FPC scores and each covariate where the weight is the estimated SFPS weight obtained by either the parametric (Para) or nonparametric (Np) method. We also computed the unweighted absolute Pearson correlation for comparison. The results are illustrated in Figure \ref{fig:data_bal}. As shown in Figure \ref{fig:data_bal}, both covariate balancing methods can always
reduce the imbalance between the FPC scores and multivariate covariate, and their improvements for the logarithm of height and its square are particularly substantial. The performances of the two methods are comparably satisfactory, although the parametric one is mostly slightly better.



To estimate the causal effect of body circumference on the VAT-to-weight ratio, we considered a functional linear outcome model which includes $X$, gender (female$=0$/male$=1$) and their interaction. To fit the outcome model which involves two unknown coefficient functions for $X$ and interaction term respectively, we followed the same truncation approach in Section \ref{sec:ce} where both coefficient functions are approximated by a finite number of eigenfunctions of $X$ and the functional linear model is accordingly approximated by a classical linear model on the corresponding FPC scores, gender and their interactions. 

Due to the importance of outcome model specification as indicated in Section \ref{sec:simu}, we 
adopted the association--variation index (AVI) proposed by \citet{SuDH17} to select eigenfunctions/FPC scores of $X$ used in the approximate outcome model.
Explicitly the AVI for the $k$th eigenfunction/FPC score is defined by $V_k=\lambda_k \beta_k^2$, 
where $\lambda_k=\text{Var}(A_k)$ and $\beta_k = \text{Cov}(A_k, Y)/\text{Var}(A_k), k \geq 1$, which takes into account both the variation of $X$ each FPC score $A_k$ explains and $A_k$'s predictability of the outcome $Y$. For each $k \geq 1$, 
$\lambda_k$ may be estimated 
by the sample variance of $\{A_{ik}: i=1, \ldots, n\}$ and $\beta_k$ by 
may be estimated by 
the ordinary least square estimate of the slope when regressing $Y$ on $A_k$. Following the procedure by \citet{SuDH17}, we first obtained an initial set of four eigenfunctions/FPC scores by setting $\text{PVE} = 0.99$ together with their estimated AVI values $\hV_k, k=1, \ldots, 4$, and then obtained their sorted AVI values $\hV_{(k)}, k=1, \ldots, 4$ in a decreasing order. Finally we selected the eigenfunction/FPC scores concomitant with the largest three AVI values such that their cumulative percentage of association--variation explained is at least $0.99$, that is, $3 = \argmin_{K \leq 4} \{\sum_{k=1}^K \hV_{(k)} / \sum_{l=1}^4 \hV_{(l)} \geq 0.99\}$. The selected eigenfunctions are the first three 
in terms of their eigenvalues as illustrated in Figure \ref{fig:efn}, and their associated FPC scores are used in the approximate outcome model. The results for fitting the approximate outcome model are given in Table \ref{tab:om}. 

\begin{table}[H]
\caption{Coefficient estimates, standard error (SE) and p-values for fitting the approximate outcome model after weighting by 
the parametric (Para) and nonparametric (Np) covariate balancing methods respectively. F-statistics for the overall model significance are also given, together with the corresponding p-values based on na\"{i}ve F-tests which ignore the uncertainty in the process of estimating SFPS weights and selecting FPCs. 
}
\centering
\label{tab:om}
\begin{tabular}{*{7}{c}}
  \multirow{2}*{}  & \multicolumn{3}{c}{Para} & \multicolumn{3}{c}{Np} \\
  \cmidrule(lr){2-4} \cmidrule(lr){5-7}
 & Estimate & SE & P-value & Estimate & SE & P-value \\ 
  \hline
Intercept & 32.94 & 2.11 & $< 10^{-6}$ & 32.40 & 2.23 & $< 10^{-6}$ \\ 
  FPC1 & -0.22 & 0.02 & $< 10^{-6}$ & -0.23 & 0.02 & $< 10^{-6}$ \\ 
  FPC2 & -0.36 & 0.07 & $1.07\times10^{-6}$ & -0.41 & 0.07 & $< 10^{-6}$ \\ 
  FPC3 & 0.02 & 0.14 & 0.8950 & 0.10 & 0.14 & 0.4550 \\ 
  gender & -17.28 & 5.99 & 0.0045 & -17.37 & 5.54 & 0.0021 \\ 
  FPC1$\times$gender & -0.05 & 0.04 & 0.1396 & -0.04 & 0.04 & 0.2337 \\ 
  FPC2$\times$gender & -0.04 & 0.15 & 0.7995 & 0.01 & 0.14 & 0.9712 \\ 
  FPC3$\times$gender & 0.04 & 0.27 & 0.8748 & -0.03 & 0.24 & 0.8869 \\ 
   \hline
  & \multicolumn{3}{c}{F-statistic=33.38, P-value$< 10^{-6}$} & \multicolumn{3}{c}{F-statistic=33.62, P-value$< 10^{-6}$} \\
  \hline
\end{tabular}
\end{table}

Table \ref{tab:om} demonstrates an overall significant causal effect of body circumference on VAT-to-weight ratio 
at the level of significance $0.05$ no matter if the covariates are balanced by the parametric or nonparametric method. It also shows that the SFPS weightings by the parametric and nonparametric balancing method lead to almost identical outcome model fitting results. 


\begin{figure}[H]
    \centering
    \includegraphics[width=12cm]{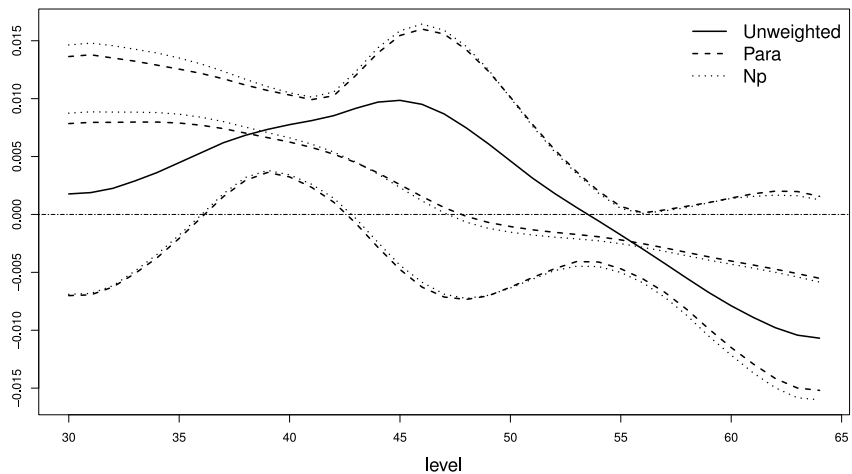}
    \caption{Estimated causal effects of body circumference on VAT-to-weight ratio for female subjects where the multivariate covariate is either unadjusted (Unweighted) or adjusted for by the parametric (Para) or nonparametric (Np) covariate balancing method. The 99\% bootstrap pointwise confidence intervals are given for the two covariate balanced casual effect estimates.
    } 
    \label{fig:data_all_ecf_f}
\end{figure}

\begin{figure}[H]
    \centering
    \includegraphics[width=12cm]{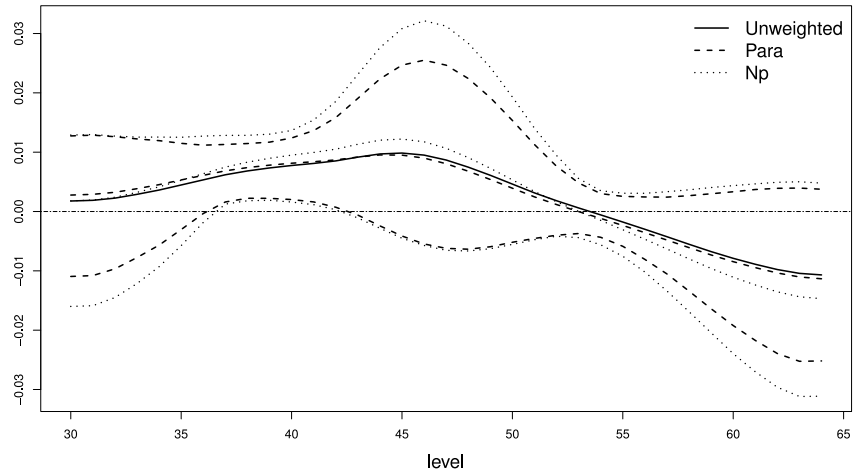}
    \caption{
    The same as Figure \ref{fig:data_all_ecf_f} except for the causal effect for males. } 
    \label{fig:data_all_ecf_m}
\end{figure}

\begin{figure}[H]
    \centering
    \includegraphics[width=12cm]{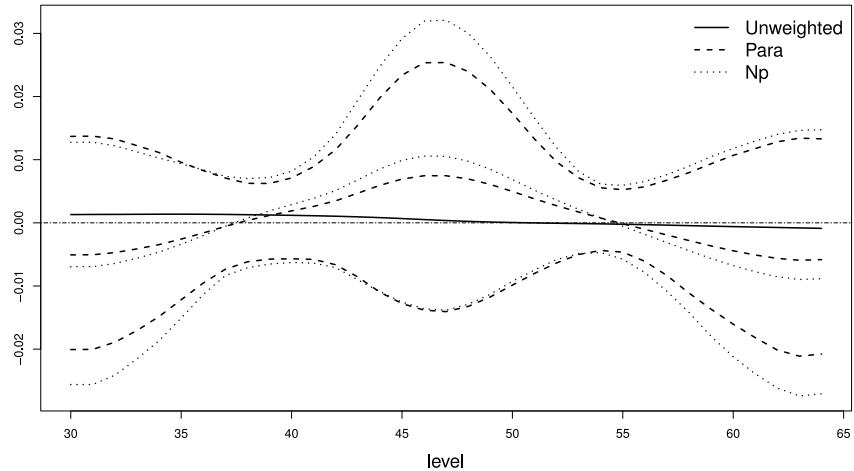}
    \caption{
    The same as Figure \ref{fig:data_all_ecf_f} except for the causal effect difference between males and females.}
    \label{fig:data_all_ecf_d}
\end{figure}

The estimated causal effects of body circumference for females, males and their difference are illustrated in Figures \ref{fig:data_all_ecf_f}, \ref{fig:data_all_ecf_m} and  \ref{fig:data_all_ecf_d} respectively, together with their 99\% pointwise confidence intervals based on $10,000$ bootstrap samples. Similar to the observation from Table \ref{tab:om}, all three figures show that the parametric and nonparametric covariate balancing methods lead to very similar causal effect estimates. They also show that the unadjusted estimates seem to be seriously biased for the causal effect for females (Figure \ref{fig:data_all_ecf_f}) and the causal effect difference between males and females (Figure \ref{fig:data_all_ecf_d}), but not for the causal effect for males (Figure \ref{fig:data_all_ecf_m}). 

Figure \ref{fig:data_all_ecf_f} shows that the causal effect of body circumference on VAT-to-weight ratio is overall significant for females no matter if the covariates are balanced by the parametric or nonparametric method. This can be validated by p-values $< 0.0001$ based on na\"{i}ve F-tests 
which ignore the uncertainty in the process of selecting FPCs and estimating SFPS weights.
The domain for the significant causal effect is between levels $36$ and $43$, 
which approximately corresponds to the body region below chest to navel, 
 and the causal effect gradually decreases from level $36$ to level $43$. Figure \ref{fig:data_all_ecf_m} shows that the causal effect of body circumference for males is also overall significant, with p-values $<0.0001$ 
by na\"{i}ve F-tests. The significant domain for the causal effect is essentially the same as that for females, but the causal effect tends to slightly increase from level $36$ to level $43$. Despite the seeming different causal effect patterns between females and males, Figure \ref{fig:data_all_ecf_d} shows no significant difference in the causal effect of body circumference between the two gender groups, with p-values 0.5274 (Para) and 0.6538 (Np) respectively by na\"{i}ve F-tests. 




\section{Discussion} \label{sec:dis}


%
%
%
%

To the best of our knowledge, this paper is the first to study the causal effect estimation of functional treatments in observational studies. Due to the lack of a probability density function for a functional variable in general, we properly define the FPS in terms of a multivariate substitute for the functional treatment, i.e., its FPC scores. We propose two covariate balancing methods to estimate the FPS weights, which are used in an outcome model to estimate the causal effect by FPS weighting. The appealing numerical performance of the proposed method in both imbalance reduction and causal effect estimation accuracy is demonstrated by a simulation study. Using the proposed method, the paper has made the first endeavor to study the causal effect of body circumference on VAT among all body-shape-based analyses of VAT. 

The proposed method may be straightforwardly generalizable to other scenarios. For example, it is almost directly applicable to handle multidimensional continuous treatments \citep[e.g.,][]{KongYW19}
and functional/categorical outcomes. 
If the covariate set consists of a functional variable, the proposed method is also applicable by including its top FPC scores in $\bC$ \citep[e.g.,][]{MiaoXZ20}. With slight modifications, the proposed method may be used to study the joint causal effect of multiple functional treatments if the FPS and SFPS are defined in terms of their joint FPC scores obtained by multivariate functional principal component analysis \citep[e.g.,][]{ChioCY14, HappG18}. 




This paper has a few limitations which are worthy of future studies. First, the proposed method relies on the FPC scores of the functional treatment. A consistent recovery of them is attainable if the functional treatment is fully observed or densely measured, 
but not if it is sparsely and irregularly measured \citep[e.g.,][]{YaoMW051}. 
A future research topic is to develop causal effect estimation methods for sparsely and irregularly observed functional treatments. Moreover, it is worthwhile to study the numerical improvements on both parametric and nonparametric methods for FPS estimation since they may lead to a poor covariate balance if a large number of FPC scores is selected or the number of covariates is moderate or large. Another interesting research direction is to study non-truncation methods for outcome model fitting, e.g., via roughness regularization.






\section*{Acknowledgements}\label{sec:ack}
The research of Xiaoke Zhang was partially supported by the USA National Science Foundation under grant DMS-1832046. We are thankful for Dr. James K. Hahn who kindly provides the VAT dataset. 

\bibliographystyle{chicago}
\bibliography{psw,mfar,nsf-xiaoke}

\end{document}